\documentclass[twocolumn,showpacs,preprintnumbers,amsmath,amssymb,letter]{revtex4-1}

\usepackage[usenames]{color}
\usepackage{graphicx}
\usepackage{bm}
\usepackage{rotate}
\usepackage{multirow}
\usepackage[normalem]{ulem}

\newcommand{\ave}[1]{\ensuremath{\langle #1 \rangle}}
\def\corr{\ave{x_i\cdot x_{i+\ell}}}
\def\corrabs{\ave{|x_i |\cdot |x_{i+\ell}|}}
\newcommand{\sgn}[1]{\ensuremath{{\rm sgn}(#1)}}
\newcommand{\abs}[1]{\ensuremath{\vert #1 \vert}}

\def\K{\ensuremath{K_{x}}}
\def\Kabs{\ensuremath{K_{|x|}}}
\def\Ksq{\ensuremath{K_{x^2}}}
\def\C{\ensuremath{C_{x}}}
\def\Cabs{\ensuremath{C_{|x|}}}
\def\Cexp{\ensuremath{C_{|x_{\rm exp}|}}}
\def\Cs{\ensuremath{C_{s}}}   
\def\Csq{\ensuremath{C_{x^2}}}
\def\sigmax{\ensuremath{\sigma_x}}
\def\sigmaabs{\ensuremath{\sigma_{|x|}}}
\def\sigmasq{\ensuremath{\sigma_{x^2}}}

\def\intinf{\int_{0}^{\infty}}
\def\intinfinf{\int_{-\infty}^{\infty}}
\def\Ak{\ensuremath{A_{k}}}
\def\medmod{\ensuremath{\sqrt{\frac{2}{\pi}}}}
\def\modmed{\ensuremath{\sqrt{\frac{\pi}{2}}}} 
\newcommand{\e}[1]{\ensuremath{\exp \left( #1 \right)}}

\begin{document}

\preprint{APS/123-QED}

\input epsf.sty

\title{Correlations in magnitude series to assess nonlinearities: application to multifractal models and heartbeat fluctuations}

\author{Pedro A. Bernaola-Galv\'an$^1$, Manuel G\'omez-Extremera$^1$, A. Ram\'on Romance$^2$ and Pedro Carpena$^1$}

\address{$^1$ Dpto. de F\'{\i}sica Aplicada II. ETSI de Telecomunicaci\'on. University of M\'alaga. 29071 M\'alaga, Spain
\\ $^2$ Dpto. de Did\'actica de la Lenguas, las Artes y el Deporte. Facultad de C.C. E.E. University of M\'alaga. 29071 M\'alaga, Spain.}

\begin{abstract}
The correlation properties of the magnitude of a time series are associated with nonlinear and multifractal properties and have been applied in a great variety of fields.  
{Here, we have obtained the analytical expression of the autocorrelation of the magnitude series ($C_{|x|}$) of a linear Gaussian noise as a function of its autocorrelation ($C_x$)}. 
For both, models and natural signals, the deviation of $C_{|x|}$ from its expectation in linear Gaussian noises can be used as an index of nonlinearity that can be applied to relatively short records and does not require the presence of scaling in the time series under study. 
In a model of artificial Gaussian multifractal signal we use this approach to analyze the relation between nonlinearity and multifractallity and show that the former implies the latter but the reverse is not true.
We also apply this approach to analyze {experimental data:} heart-beat records during rest and moderate exercise. For each individual subject, we observe higher nonlinearities during rest. This behavior is also achieved on average for the analyzed set of 10 semiprofessional soccer players. This result agrees with the fact that other measures of complexity are dramatically reduced during exercise and can shed light on its relationship with the withdrawal of parasympathetic tone and/or the activation of sympathetic activity during physical activity.
\end{abstract}

\maketitle

\section{Introduction}

In the field of time series analysis, the concept of nonlinearity can be interpreted in different ways \cite{Yosi_Geo}. An intuitive definition is that nonlinear time series are those generated by nonlinear dynamic equations, i.e. the values of the series depend on time, or on other values of the series, according to nonlinear expressions --- squares, logarithms, trigonometric functions, etc. But usually we do not have prior information about this dependence, in fact, in most of the cases the goal is nothing but finding such dynamic equations. Nonlinearity is also frequently defined in terms of the autocorrelation function: a time series is nonlinear when there is dependence between the values of the series at different positions even though its autocorrelation vanishes. 
 Although a bit more complicated, the definition of of Schreiber and Schmitz \cite{Schreiber2000} is quite suitable for practical purposes. According to this definition, a time series is linear when its Fourier phases are random, i.e. the series of phases of its Fourier transform is a random number uniformly distributed in the interval $[-\pi,\pi]$. Thus, the presence of nonlinear correlations in a time series can be assessed by means of surrogate data tests: (i)  Given a time series, compute its Fourier transform, randomize its Fourier phases and transform it back. The resulting surrogated series preserves the distribution of the data and the linear correlations because its power-spectrum remains unchanged \cite{Schreiber2000}.  (ii) Some relevant statistics is evaluated in the original as well as in the surrogated signal and, if there is a statistically significant difference
between  both signals,  it means that the original Fourier phases were not random and thus, the original signal was nonlinear, i.e. the null hypothesis of linearity can be rejected. Sometimes instead of accepting or rejecting the null hypothesis, the goal is simply to compare the degree of nonlinearity of two different time series (e.g. records obtained under different physiological conditions) and the value of the statistics is directly used as a measure of nonlinearity.

The autocorrelations in the magnitude series is also a good indicative of the presence of nonlinear correlations. For a given time series $\{y_i\}$, $i=1,...,N$, its magnitude series (sometimes also called volatility) is usually defined as the absolute value of the series increments:
\begin{equation}
    \abs{x_i}=\abs{y_{i+1}-y_i}.
    \label{def_mag}
\end{equation}
It is defined as the magnitude of the increments rather than the magnitude of the series itself because in most cases the series of increments is fairly stationary while the original series is not. {Apart from its utility in revealing nonlinear properties, the magnitude series together with the sign series (magnitude-sign analysis \cite{Yosi_PRL})  provides complementary information about the original series:
the magnitude measures how big the changes are and the sign indicates their direction.}

{Once obtained the magnitude series, the standard procedure to quantity its correlations is the use of the Detrended Fluctuation Analysis (DFA)}. In brief, the DFA method obtains the root mean square fluctuations of the series around the local trend $F_d(\ell)$ in all windows of a given size $\ell$ and repeats the procedure for different window sizes. Scaling is present when 
\begin{equation}
 F_d(\ell) \propto \ell^\alpha.
 \label{DFA}
\end{equation}
Typically, $\alpha$ is estimated as the slope of a linear fitting of $\log(F_d)$ vs. $\log(\ell)$. The exponent $\alpha$ quantifies the strength of the correlations present in the time series and is also related to the power spectrum exponent $\beta$ and the autocorrelation function exponent $\gamma$ \cite{allegrini,rangarajan}.
The scaling analysis of the magnitude series,  was first introduced to study nonlinearities in heart-beat fluctuations  \cite{Yosi_PRL} but since then, examples of quantifying nonlinearity using the DFA exponent of the magnitude series can be found in many other fields such as Fluid Dynamics \cite{fluid}, Geophysical \cite{geo01,geo02,Yosi_Geo} and Economical time series \cite{economy}. 

The scaling exponent {of the magnitude fluctuations} is easy to compute and is also related to the width of the multifractal spectrum \cite{Yosi_volatility,manolo}, another quantity also frequently used to unveil the nonlinear properties of a signal \cite{Plamen1999}.

Nevertheless, this approach shows three main drawbacks: 

\begin{itemize}
\item[(i)] In order to properly define the scaling exponent $\alpha$, $F_d(\ell)$ vs. $\ell$ must show a good fit to a power-law, which is not the case in many natural series. Also, the interpretation of crossovers in $F_d(\ell)$ vs. $\ell$ as a signature of the existence of regions with different scaling has been recently challenged. In particular, it has been shown that the evaluation of short-range scaling exponent ($\alpha_1$), a quantity widely used in heart rate analysis  \cite{Pena09}, could be affected by spurious results \cite{Holl15} and that $\alpha_1$ is strongly biased by the breathing frequency \cite{Pandelis09}. Without judging the validity of these criticisms, the truth is that some results obtained with $\alpha_1$ are contradictory \cite{Sandercock06}. These problems affect DFA in general as a technique to evaluate scaling exponents. 
\item[(ii)]   Furthermore, particularizing to the evaluation of the scaling exponent of the magnitude series, we have shown very recently that in some situations DFA does not properly detect the correlations and assigns uncorrelated behavior to correlated magnitude series \cite{CarpenaDFA}.
\item[(iii)] It is assumed implicitly that the presence of correlations in the magnitude series is a signature of nonlinearity but, as we show later, even for a linear time series, $\Cabs > 0$ when $\C \neq 0$.
\end{itemize}


For these reasons we propose here a different approach: We consider a linear Gaussian noise $\{x_i\}$,  i.e. a time series whose values follow a Gaussian distribution and its Fourier phases are random, with correlations given by $\C$  and obtain analytically the correlations $\Cabs$ of the magnitude series $\{\abs{x_i}\}$ which result to depend only on $\C$. Note that $\Cabs$ are the magnitude correlations expected in purely linear noises. When analyzing an experimental time series $\{x_{{\rm exp}}(i)\}$, the deviation of the correlation of its magnitude $\Cexp$ with respect to the linear expectation $\Cabs$ is then a good signature of nonlinear correlations.  Taking into account that natural data does not always follow Gaussian distributions, prior to the computation of the magnitude correlations, we transform the distribution of the data to a normal distribution with zero mean and unit standard deviation, ${\cal N}(0,1)$.

This article is organized as follows: {Motivated by the fact  that in most of the examples cited above, correlated non-stationary natural series are modeled as fractional Brownian motions (fBm), and thus their stationary increments as fractional Gaussian noises (fGn),} in section \ref{section_cmag} we obtain the analytical expression of the autocorrelation of magnitude series of a linear Gaussian noise as a function of its autocorrelation as well as a quadratic approximation. We also obtain the corresponding expression for the series of squares, $\{x_i^2\}$,  which is sometimes used to study nonlinear correlations (Sec. \ref{section_csq}) and discuss about the autocorrelation of the sign series and its relation to the autocorrelation of the magnitude series (Sec. \ref{section_csign}). In section  \ref{section_nonlinear} we explore the nonlinear properties of artificial series generated with a model that produces Gaussian noises with multifractal properties and in section \ref{section_natural_signals}, as an example of their utility, we apply the relations derived here to the study of heart beat time series during rest and moderate exercise. Section \ref{conclusions} concludes the paper.

\section{Autocorrelation of magnitude series}

\label{section_cmag}

Given a time series $\lbrace y_i \rbrace$, with its corresponding series of increments $\lbrace x_i \rbrace$, our aim is to obtain the autocorrelation $\Cabs(\ell)$  of its magnitude series $\lbrace | x_i | \rbrace$ (\ref{def_mag})  as a function of the autocorrelation of the  series of increments  $\C(\ell)$ provided that $\lbrace x_i \rbrace$ is  a linear Gaussian noise, i.e. all $x_i \sim {\cal N}(0,1) $  and that only linear correlations are present in the series. Thus, the autocorrelation function at distance $\ell$ is given by:
\begin{equation}
   \C(\ell) = \frac{\corr-\ave{x_i} \ave{x_{i+\ell}}}{\sigmax^{2}}=\corr,
\end{equation}
where \ave{\cdot} denotes average over the series and $\sigmax^{2}$ is the variance of the series.

Under the assumption of $x_i \sim {\cal N}(0,1)$ the autocorrelation coincides with the autocovariance
\begin{equation}
\K(\ell) \equiv \corr - \ave{x_i}\ave{x_{i+\ell}} = \C(\ell).
\end{equation}
On the other hand, for the magnitude series we have:
\begin{eqnarray}
   \sigmaabs^{2}&=& 1- \frac{2}{\pi},
\end{eqnarray}
and we can write for its autocorrelation:
\begin{eqnarray}
  \Cabs(\ell) &=& \frac{\corrabs-\ave{|x_i|}\ave{|x_{i+\ell}|}}{\sigmaabs^{2}}  \nonumber \\
  &=& \frac{\pi \Kabs(\ell)}{\pi - 2} ,
  \label{cmag1}
\end{eqnarray}

Taking into account that $x_i$ and $x_{i+\ell}$ are two linearly correlated Gaussian random variables,   
the autocovariance of the magnitude series, $\Kabs(\ell)$, can be expressed as a function of $\K(\ell)$ according to Eq. (\ref{cov_mag}) in appendix \ref{append_mod}:

\begin{equation}
 \Kabs= \frac{2}{\pi} \left[ \sqrt{1-\K^2} + \K\arcsin {\K} -1 \right],
\end{equation}
and replacing in (\ref{cmag1}):
\begin{equation}
  \Cabs = \frac{2 \left [\C \arcsin\C-1 + \sqrt{1-\C^2}  \right ] }{\pi-2}
  \label{cmag2}
\end{equation}

\begin{figure}
\centering
\includegraphics[width=8cm]{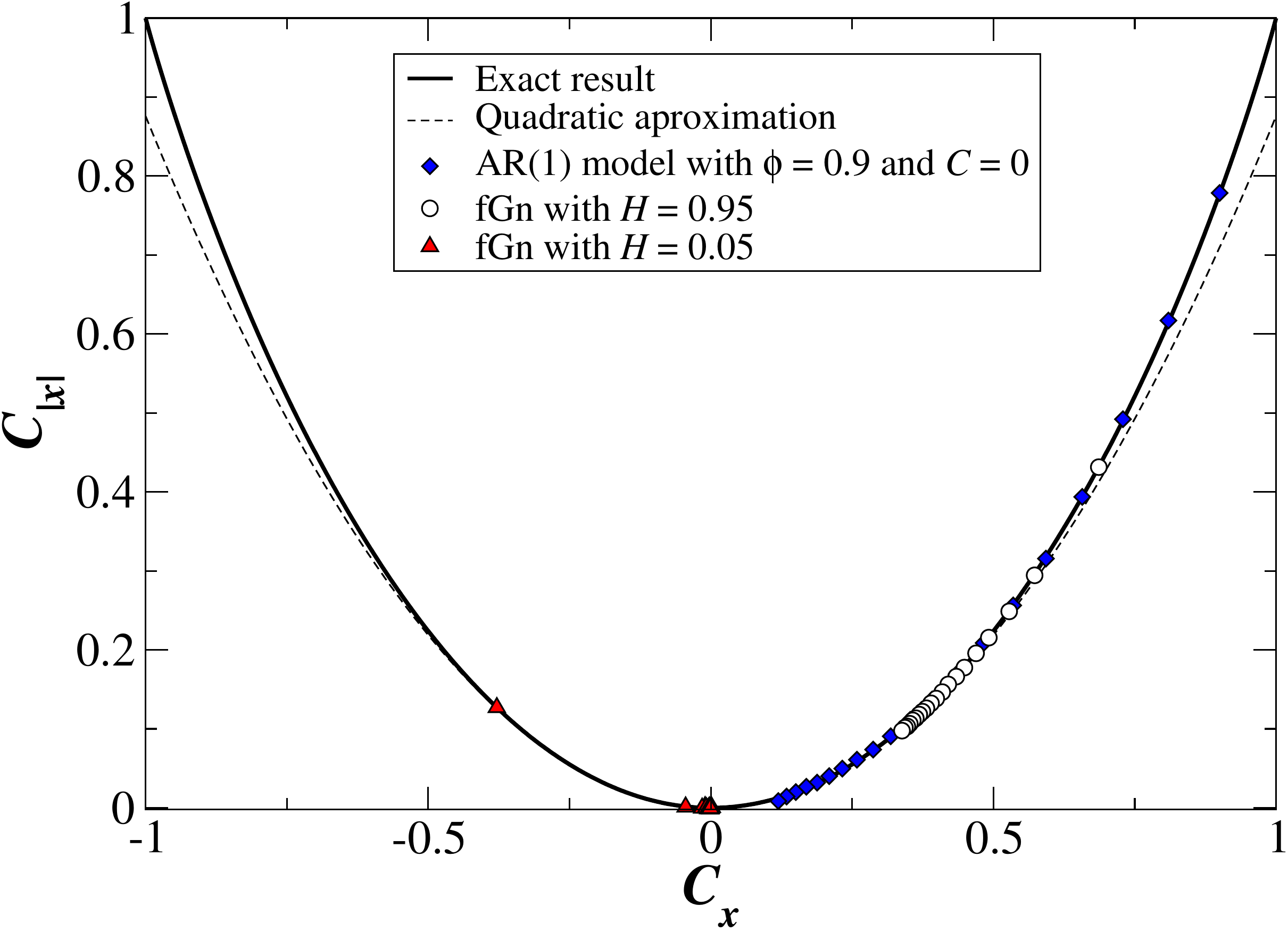}
\caption{(Color online) Autocorrelation of the magnitude series $\Cabs$ as a function of the autocorrelation of the series $\C$. Solid line corresponds to the exact expression given by Eq. (\ref{cmag2}) and dashed line to its quadratic approximation given by Eq. (\ref{cmag_aprox}). The symbols correspond to the autocorrelation at distances $\ell = 1,...,20$ for several artificial series generated with linear Gaussian models: diamonds, autorregresive AR(1) model $x_i=c+\phi  x_{i-1}+\varepsilon_i$, with $\phi = 0.9$, $c=0$ and $\{\varepsilon_i\}$ a white noise; circles, fractional Gaussian noise (fGn) with Hurst exponent $H = 0.95$ and triangles fGn with $H=0.05$. While the two first models (diamonds and circles) generate highly correlated series the last one (triangles) leads to an anticorrelated series. However, in all cases the correlations of the magnitude series are positive.}
\label{fig_parabola}
\end{figure}

It is easy to check that (\ref{cmag2}) is an even and positive function which implies that the magnitude of a linear Gaussian noise cannot be anticorrelated (Fig. \ref{fig_parabola}).

If we consider small values of $\C$,
Eq.  (\ref{cmag2}) can be approximated by a Taylor expansion and obtain:

\begin{equation}
 \Cabs =  \frac{1}{\pi-2} \C^2  + {\cal O}(\C^4) 
 \label{cmag_aprox}
\end{equation}

Thus, for small values, $\Cabs$ behaves essentially as the square of $\C$.
In fact, the error of (\ref{cmag_aprox}) is around $2\%$ for $\C=0.5$ which makes this approximation virtually correct for most real data. In figure \ref{fig_parabola} we plot Eq. (\ref{cmag2}), its quadratic approximation Eq. (\ref{cmag_aprox}) as well as several examples of artificial series created with Gaussian linear models.

This result is especially interesting when studying the scaling behavior of series with power-law correlations that have been found in a great variety of complex systems.  We can characterize these series by their power spectral exponent $\beta$ because most methods of generating power-law correlated Gaussian noises consist in the generation of series with $1/f^\beta$ decay in their power spectrum with  $-1<\beta<1$ (e.g.  \cite{multifractal,Makse}). In particular, these methods are widely used to generate approximate fractional Gaussian noises (fGn) which are indeed linear Gaussian noises  whose autocorrelation function decays asymptotically as a power law \cite{Beran}:
\begin{equation}
\C(\ell) \simeq \frac{(1-\gamma)(2-\gamma)}{2\ell^\gamma} \propto \frac{\sgn{1-\gamma}}{\ell^\gamma} 
\label{corrpw}
\end{equation}
where $\gamma = 1-\beta$. It is also quite common to characterize the fGns by their Hurst exponent ($H$) which is related to both $\beta$ and $\gamma$ by:
\begin{equation}
 H= \frac{\beta+1}{2}=\frac{2-\gamma}{2}
 \label{Hurst}
\end{equation}

For stationary time series ($0<H<1$), the Hurst exponent also coincides with the DFA exponent $\alpha$ (\ref{DFA}).
Note that the term $\sgn{1-\gamma}$ in the numerator of (\ref{corrpw}) vanishes  for $\beta=0$ ($H=0.5$, white noise) thus leading to an uncorrelated random noise. For $\beta>0$ ($H>0.5$) the series is long-range correlated, also known as having  ``long memory'' \cite{rangarajan,Beran}, in the sense that its autocorrelation decays very slow with exponent $\gamma < 1$. In fact  $\sum_{\ell=1}^{L}\C(\ell)$ diverges as $L \rightarrow \infty$.
Likewise for $\beta<0$ ($H<0.5$), i.e. $\gamma>1$, $\C$ is negative and the series is anticorrelated. In this situation, although the autocorrelation also decays as a power law, we cannot properly speak about {\sl long-range} anti-correlations because they decay relatively fast, in the sense that now the autocorrelation function is summable. Another conclusion drawn from (\ref{corrpw}) is that we cannot obtain linear Gaussian noises with positive autocorrelation functions decaying faster than $1/\ell$.
 

\begin{figure}
\centering
\includegraphics[width=8cm]{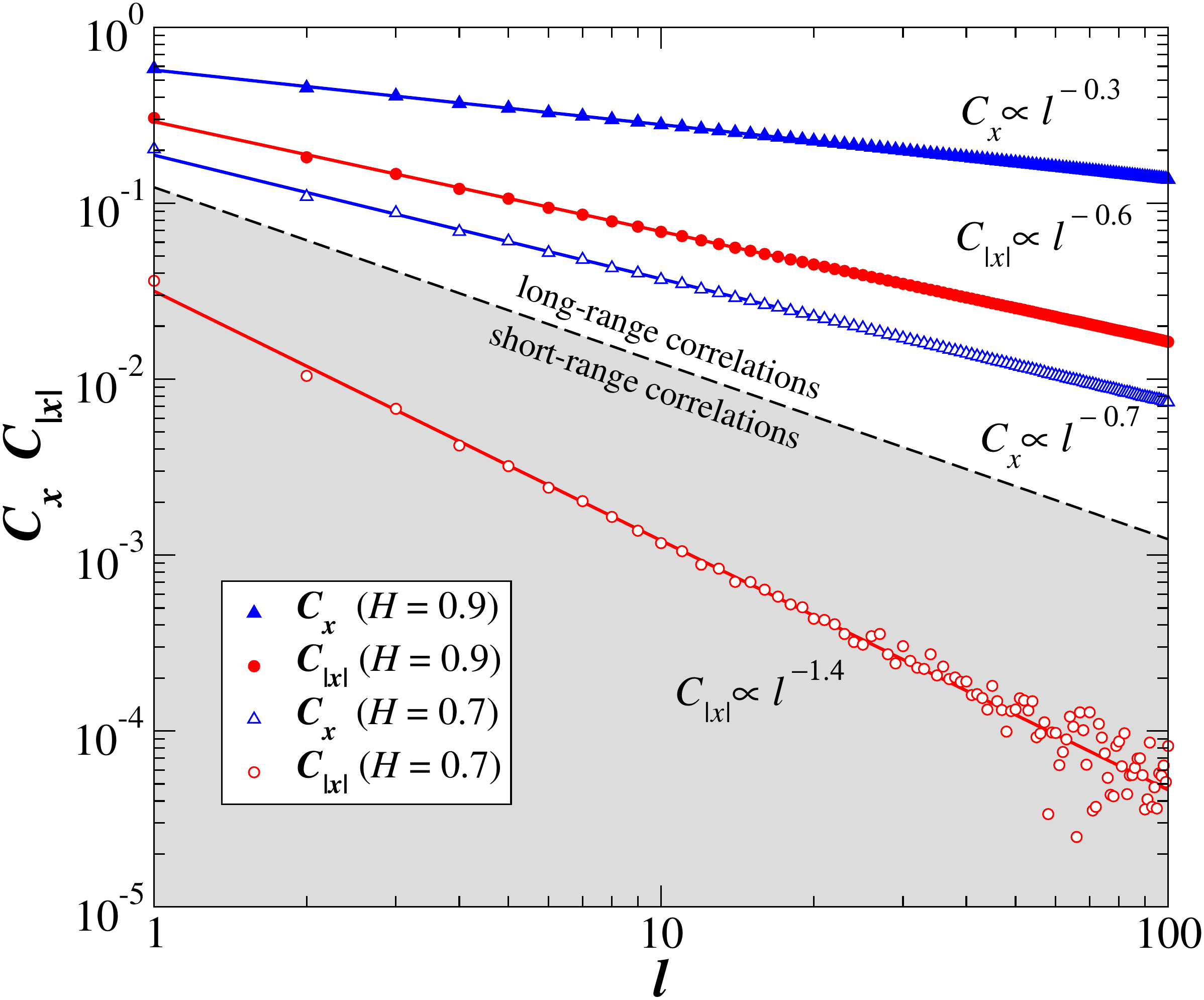}
\caption{(Color online) Autocorrelation function $\C$ (triangles) and autocorrelation function of magnitude series $\Cabs$ (circles) as a function of distance $\ell$ for two examples of fractional Gaussian noises (fGn): $H=0.9>0.75$ (full symbols) and $H=0.7<0.75$ (open symbols). We generate series of approximate fGns of length $2^{24}\simeq 1.6\times 10^7$    using the Fourier Filtering Method \cite{Makse}. To avoid statistical fluctuations we average over an ensemble of 100 realizations. The dashed line corresponds to $1/\ell$ which is the boundary between the regions of long and short-range correlations (see text).
As expected, $\Cabs$ decays with an exponent double that of $\C$ and in both cases the magnitude series are positively correlated, but for $H=0.7 < 0.75$  $\Cabs$ decays faster than $1/\ell$ and lies in the short-range correlations region. Note also that in this case and due to both the fast decay of $\Cabs$ and its small values at $ \ell=1$, a noisy behavior is reached relatively soon ($\ell<100$) even for long series. This makes difficult the detection of power-law behaviors in this region when dealing with real data.}
\label{fig_corrpw}
\end{figure}

We obtain from  (\ref{cmag_aprox})  that the autocorrelation of the magnitude series of a fGn also decays as a power law with exponent $2\gamma$ and is always positive, even for $H<0.5$ when the fGn is anticorrelated:
\begin{equation}
  \Cabs \propto \frac{1}{\ell^{2\gamma}}
  \label{corrabspw}
\end{equation}

Nevertheless, we must distinguish two  different situations:
\begin{itemize}
\item[(i)] $H>0.75$. Here $2\gamma<1$ and $\Cabs$ decays slower than $1/\ell$ thus leading to long-range power-law correlations in the magnitude series.
\item[(ii)]  $H<0.75$. Now $2\gamma>1$ and $\Cabs$, although still being positive and following a power-law, decays very fast. For example, in Fig. \ref{fig_corrpw} we can see that for $H=0.7$, $\Cabs$ reaches the background noise level for relatively short scales ($\ell < 100$) even for a time series as long as $2^{24}\simeq 1.6\times 10^7$.

Indeed, the methods quantifying correlations by means of the study of fluctuations fail to detect the power-law correlations present in magnitude series for $H<0.75$. For example, two widely used techniques like Fluctuation Analysis (FA) or Detrended Fluctuation Analysis (DFA) \cite{Peng94} wrongly classify as ``white noise'' the magnitude series of Gaussian noises with $H<0.75$ \cite{manolo,Yosi_volatility}  despite being true only for $H=0.5$ \cite{CarpenaDFA}.
\end{itemize}

\subsection{Relation with the autocorrelation of square series}

\label{section_csq}

For simplicity, sometimes the autocorrelation of square series, $\{x_i^2\}$, is studied instead of the magnitude series, i.e.:
\begin{equation}
\Csq(\ell)=
\frac{\ave{x_i^2\cdot x_{i+\ell}^2}-\ave{x_i^2}\ave{x_{i+\ell}^2}}{\sigmasq^2}
\end{equation}
Indeed, it has been shown numerically that the scaling properties of the correlations of both series are quite similar \cite{Yosi_volatility}. Below we justify analytically this similarity.

As we did for the magnitude series, first we obtain the autocovariance of the square series, \Ksq, as a function of the autocovariance of the series, $\K$ (Appendix \ref{append_square}):
\begin{equation}
\Ksq(\ell) \equiv \ave{x_i^2\cdot x_{i+\ell}^2} - \ave{x_i^2}\ave{x_{i+\ell}^2} = 2 \K(\ell)^2.
\end{equation}

Taking into account that $x_i \sim {\cal N}(0,1)$ and thus $\sigmasq=\sqrt{2}$, we obtain:
\begin{equation}
\Csq = \C^2
\label{csq}
\end{equation}
Which obviously implies that the squares of a linear Gaussian noise, just as the magnitude, cannot be anticorrelated.

Eq. (\ref{csq}) also justifies the fact that for power-law correlated series, $\Cabs$ and $\Csq$ scale asymptotically with the same exponent: for long enough values of $\ell$ we have $\C\ll1$ and thus the approximation (\ref{cmag_aprox}) is valid, leading to $\Cabs \propto \Csq$.

\begin{figure}
\centering
\includegraphics[width=8cm]{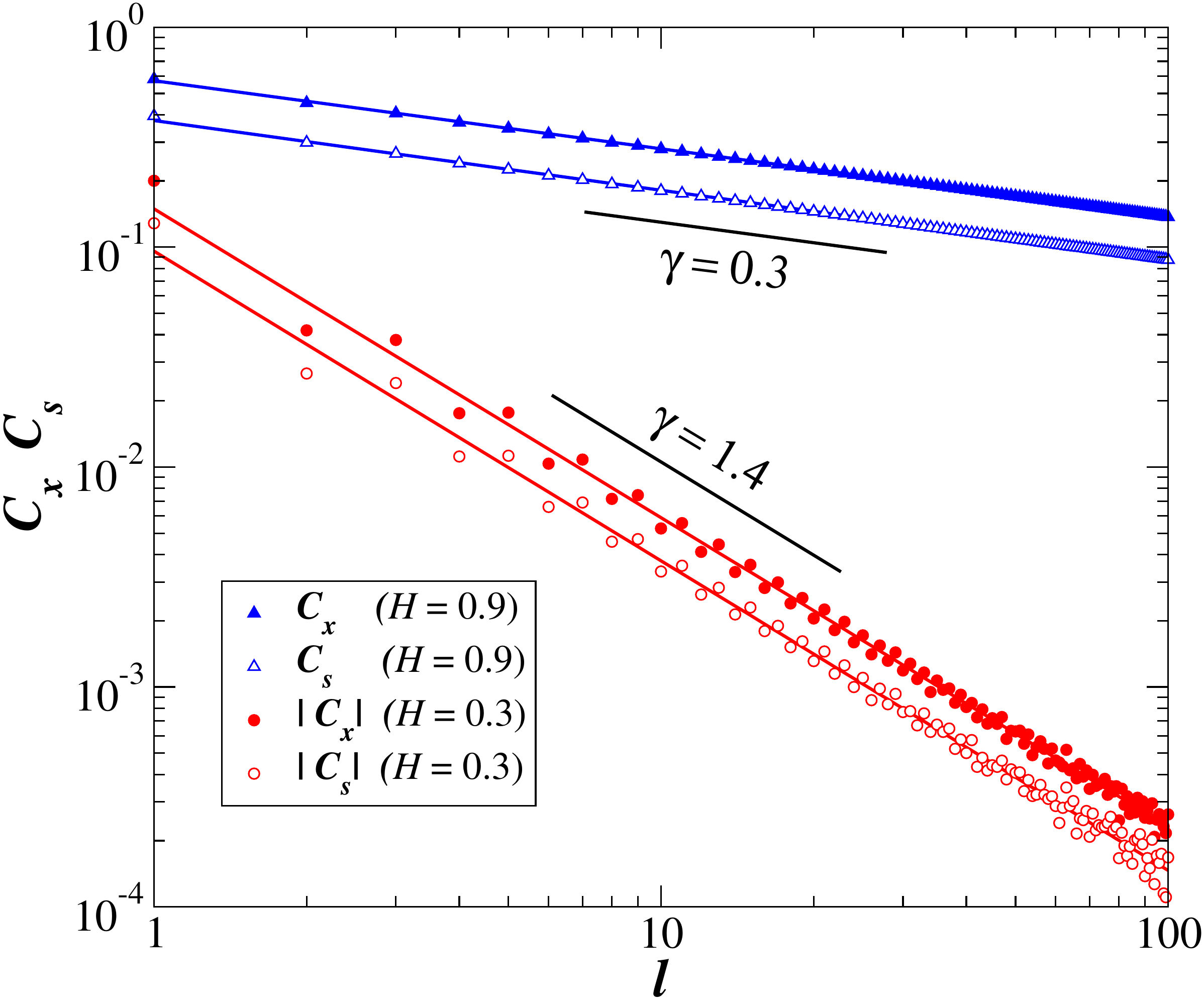}
\caption{(Color online) 
Autocorrelation function $\C$ (full symbols) and autocorrelation function of sign series $\Cs$ (open symbols) as a function of distance $\ell$ for two examples of fractional Gaussian noises (fGn): correlated $H=0.9 >0.5$ (triangles) and anticorrelated $H = 0.3<0.5$ (circles). In this case we plot \abs{\C} and \abs{\Cs} to allow the representation in double logarithmic scale. We generate series of approximate fGns of length $2^{24}\simeq 1.6\times 10^7$    using the Fourier Filtering Method \cite{Makse}. To avoid statistical fluctuations we average over an ensemble of 100 realizations.
As can be expected from (\ref{csig_approx}), $\C$ and $\Cs$ decay with the same exponent. Note that for the anticorrelated series ($H=0.3$) the anti-correlations decay very fast ($\gamma = 1.4$) and an oscillatory behavior can be observed. This oscillation is amplified for $H<0 $ (not shown).} 
\label{fig_corrpw_sgn}
\end{figure}

\subsection{Relation with the autocorrelation of the sign series}
\label{section_csign}
Apart from its relevance in the study of nonlinear correlations, the magnitude series together with the sign series (defined below) provide complementary information about the original series $\{y_i\}$: while the magnitude measures how big the changes are, the sign indicates their direction. Sign series are also relevant for the study of first-passage time in correlated processes \cite{Conchita12}. Below we obtain a relation between $\Cabs$ and the autocorrelation of the sign series, $\Cs$.  

Given a time series $\lbrace x_i\rbrace$, the series $\lbrace \sgn{x_i} \rbrace$ is defined by:
\begin{equation}
    \sgn{x} = \left \{ 
      \begin{array}{rll}
        -1 &{\rm if}& x < 0 \\
        0  &{\rm if}& x = 0\\ 
         1 &{\rm if}& x > 0
      \end{array}
    \right.
\end{equation}
If the series of increments  $\lbrace x_i\rbrace$ is a linear Gaussian noise, Apostolov et al. \cite{signum_corr} have shown  that the  autocorrelation of the sign series $\Cs(\ell)$ can be expressed in terms of the autocorrelation $\C(\ell)$ by: 
\begin{equation}
  \C = \sin \left(\frac{\pi}{2} \Cs\right)
  \label{csig}  
\end{equation}
Again, we can also obtain an approximation for small values of $\C(\ell)$:
\begin{equation}
 \Cs = \frac{2}{\pi}\C + {\cal O}(\C^3)
 \label{csig_approx}
\end{equation}
which implies that, if $\C$ is a power law, $\Cs$  scales asymptotically with the same exponent as  $\C$. In particular, this result holds for fGns (Fig. \ref{fig_corrpw_sgn}).

In addition and, taking into account that $-1\leq \Cs \leq 1$, from here it is clear that $\C$ and $\Cs$ always have  the same sign and thus, the sign series will be correlated where $\{x_i\}$ is correlated and anticorrelated where $\{x_i\}$ is anticorrelated (Fig.~\ref{fig_corrpw_sgn}).

Equation (\ref{csig}), together with (\ref{cmag2}) allows us to express the autocorrelation of the magnitude series as a function of the autocorrelation of the corresponding sign series:
\begin{equation}
 \Cabs= \frac{2}{\pi-2} \left[ \frac{\pi}{2}\Cs\sin\left(\frac{\pi}{2}\Cs\right)+\cos\left(\frac{\pi}{2}\Cs\right) -1 \right]
 \label{cmag_sig}
\end{equation}

\section{Example of  a nonlinear model}

\label{section_nonlinear}

Up to now we have only shown examples of linear Gaussian signals for which the derived relations among $\C$, $\Cabs$ and $\Cs$ (Eqs. (\ref{cmag2}), (\ref{csig}) and (\ref{cmag_sig}))  must hold. 
Nevertheless, if we consider nonlinear Gaussian signals, i.e. signals that, despite having a Gaussian distribution have nonrandom Fourier phases,  these relations are no longer valid and the deviation from these equations can be used as a signature of nonlinearity.   For example, if $\Cabs(\ell)\neq0$ and $\C(\ell)=0$, i.e. eq. (\ref{cmag2}) does not hold,  two values of the signal at distance $\ell$ are not linearly correlated ($\C(\ell)=0$) but they are not independent because $\Cabs(\ell)\neq0$ and thus, the signal is nonlinear according to one of the definitions given in the introduction.

 Here it is important to stress that these equations are valid for each individual value of the autocorrelation function and the possible deviations from nonlinearity can be observed without the assumption of any kind of scaling or power-law behavior in the signal.

We concentrate here on equation (\ref{cmag2}) because the correlations in the series of magnitudes  have been related to the presence of nonlinear correlations and multifractal structure \cite{Yosi_PRL,Yosi_volatility,manolo}. To show the effect of nonlinearities we generate artificial series using a simple method proposed by Kalisky et al. \cite{Yosi_volatility} which is able to generate multifractal Gaussian noises just by multiplying the sign and the magnitude of two independent linear Gaussian noises. Despite its simplicity, this method is able to independently control both the linear correlations of the signal and its multifractal spectrum width --- see also  \cite{manolo} for a systematic exploration of the method.

In brief this procedure, {\sl composition method} from now on, works as follows:
\begin{itemize}
\item[(i)] Obtain the magnitude series of a ${\cal N}(0,1)$ fGn  $\lbrace x_{{\rm mag}}(i) \rbrace$,   with Hurst exponent $H_1$ and the sign series of another ${\cal N}(0,1)$ fGn  $\lbrace x_{{\rm sign}}(i) \rbrace$,   with Hurst exponent $H_2$, where $i=1,...,N$, being  $N$ the size of the series.
\item[(ii)] Obtain the composed series as the product of the magnitude and sign series:
\begin{equation}
   x_{\rm comp}(i)=x_{\rm mod} (i) \cdot x_{\rm sign}(i)
\end{equation}
 for $i=1,...,N$.
\end{itemize}

The resulting series $\{x_{\rm comp}(i)\}$ is Gaussian by construction but it presents nonlinear correlations and Eq. (\ref{cmag2}) is not fulfilled. Instead, it can be shown that its  autocorrelation function is given by  \cite{manolo}:
\begin{equation}
  \C(\ell)=\Cs(\ell)\frac{(\pi-2)\Cabs(\ell)+2}{\pi}
  \label{cprod}
\end{equation}
where obviously $\Cabs$ and $\Cs$ coincide with the autocorrelation functions of the magnitude of the fGn with $H_1$ and the sign of the fGn with $H_2$ respectively. Note that, although $\C$ is not exactly a power law, it decays asymptotically as $1/\ell^{2-2H_2}$, i.e. the autocorrelation of the composed series decays asymptotically with the same exponent as the autocorrelation of the fGn used to obtain the sign series. Indeed, just take into account approximations (\ref{cmag_aprox}) for $\Cabs$ and (\ref{csig_approx}) for $\Cs$ and the asymptotic expression for the autocorrelation of a fGn (\ref{corrpw}) to obtain:
\begin{equation}
  \C(\ell) \simeq \frac{2H_2(2H_2-1)}{\pi^2 \ell^{2-2H_2}}\left[ \frac{H_1^2(2H_1-1)^2}{\ell^{4-4H_1}} +2 \right ].
\end{equation}
For $0<H_1,H_2 < 1$ the second summand is the leading one and we get asymptotically $\C(\ell)\propto1/\ell^{2-2H_2}$. In that sense we say that the linear correlations of the composed series are {\sl controlled} by the sign \cite{manolo}.


\begin{figure}
\centering
\includegraphics[width=8cm]{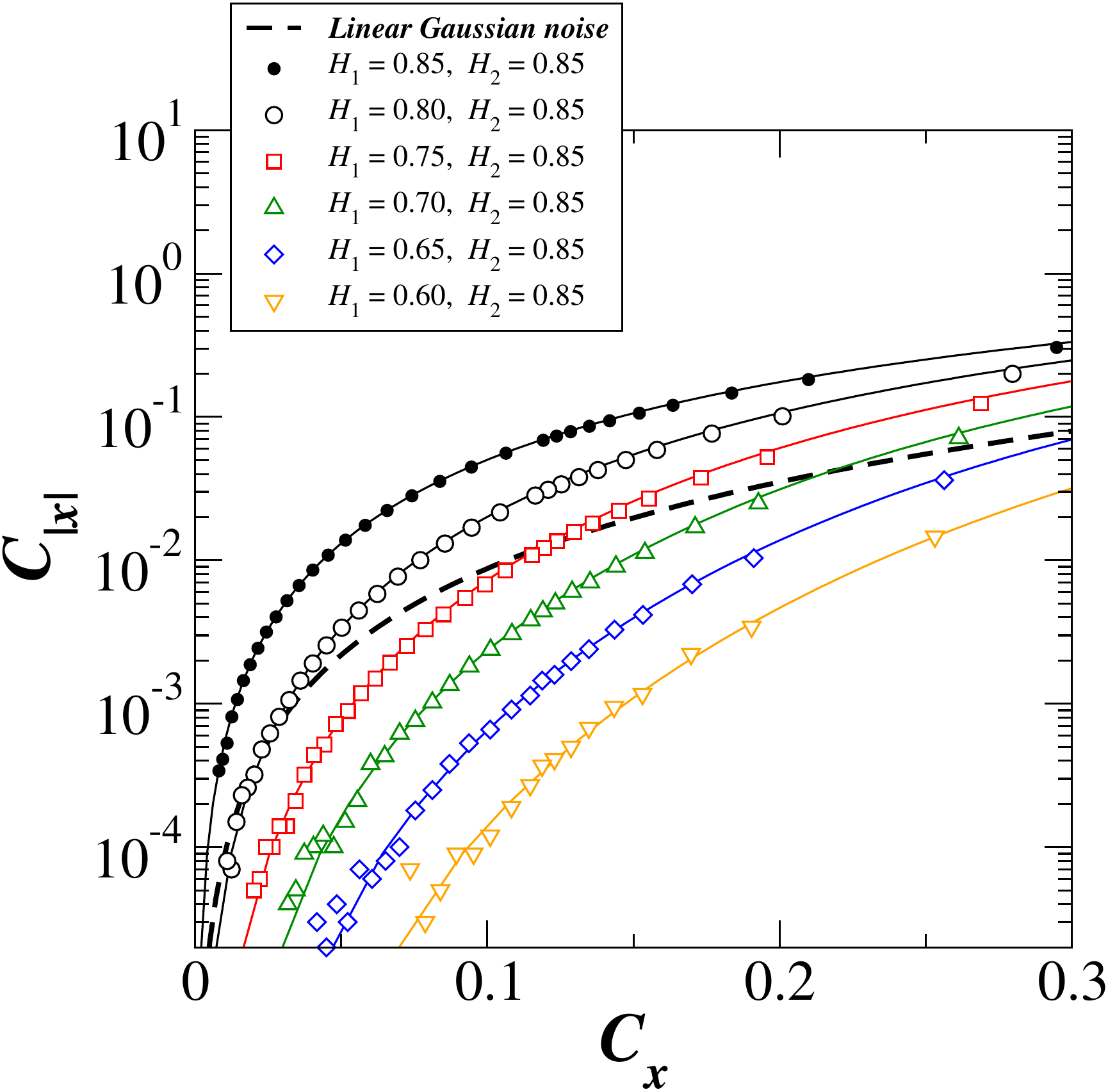}
\caption{(Color online) 
Autocorrelation of the magnitude series $\Cabs$ as a function of the autocorrelation of the series $\C$ for nonlinear series generated using the {\sl composition method} \cite{manolo} by multiplying the series of magnitudes and signs of two independent fGns (see text). Eq. (\ref{cmag2}) (dashed line) has been included as a reference for the linear Gaussian noise. Solid lines correspond to the theoretical curves obtained using (\ref{cprod}) and assuming that the fGns verify the exact asymptotic formula for their autocorrelation function (\ref{corrpw}). Eq. (\ref{cprod}) is exact, the observed deviations from these curves are due  to the fact that the series generated by means of the Fourier filtering method are approximate fGns, the expressions used for the autocorrelation are only valid asymptotically and also due to the statistical fluctuations (especially for small values of $\Cabs$).}
\label{fig_nonlinear}
\end{figure}

In Fig. \ref{fig_nonlinear} we show $\Cabs$ vs. $\C$ for several examples of nonlinear series generated by means of the composition method. For all the series shown $H_2=0.85$ and thus, all of them have the same scaling behavior for the linear correlations, nevertheless the different values of $H_1$ lead to different degrees of nonlinearity according to the deviation of $\Cabs$ from the linear expectation (dashed line in Fig. \ref{fig_nonlinear}). Note that, no matter the value of $H_1$, in all cases we observe a deviation from linearity. For smaller values of $H_1$ this deviation is more 
evident at small $\C$ (longer scales) while for large $H_1$ it appears mainly at great $\C$ (small scales). 

This means that the uncoupling of magnitude and sign (i.e. the magnitude of the changes is independent of its direction) always leads to a nonlinear behavior or, conversely, in a linear Gaussian signal magnitude and sign are not independent but coupled in a specific way that leads to the behavior described by Eq. (\ref{cmag_sig}). In general, for natural signals where magnitude and sign are neither independent nor {\sl Gaussianly coupled}, plots of $\Cabs$ vs. $\C$ can be of great utility to shed light about the way in which the magnitude of the changes is related to its direction, i.e. the magnitude and sign coupling.

\subsection{Nonlinearity and multifractality}
Multifractality and nonlinearity are two concepts that usually go together. Indeed, the width of the multifractal spectrum is considered to be linked to the degree of nonlinearity of the signal \cite{Parisi85,Badin16} and the finding of multifractal properties is usually associated with complex nonlinear interactions in the systems under study. Nevertheless, although related concepts, multifractality and nonlinearity describe the properties of the signal from different points of view \cite{Tang2015}.

The nonlinear signals generated by means of the composition method described above are a good example to show that nonlinearity is not always related to multifractality. This method was originally developed \cite{Yosi_volatility} to generate Gaussian signals with multifractal properties, in fact, it has been shown that the width of the multifractal spectrum grows linearly with the Hurst exponent $H_1$ of the signal used to obtain the magnitude series
for $H_1 > 0.75$ and when $H_1<0.75$ the width of the multifractal spectrum almost vanishes \cite{manolo}. Nevertheless, we show here (Fig. \ref{fig_nonlinear}) that for all values of  $H_1$ (including the white noise for which $H_1=0.5$) the composed signal is clearly nonlinear, despite having an almost zero mutifractal spectrum width.
Here, it is important to point out that the region where multifractal detrending techniques give null multifractal width ($H_1<0.75)$ \cite{manolo} coincides with the region where $\Cabs$ lies below the linear expectation, at least in the region where the power law fits are carried out ($\ell > 4$). According to this, we can say that for this model multifractality is a signature of nonlinearity only when the  autocorrelations in the magnitude are larger than expected in a linear model. In the opposite situation, the time series is indeed nonlinear but the multifractal analysis will not reveal it.

 \begin{figure}
 \centering
 \includegraphics[width=8.7cm]{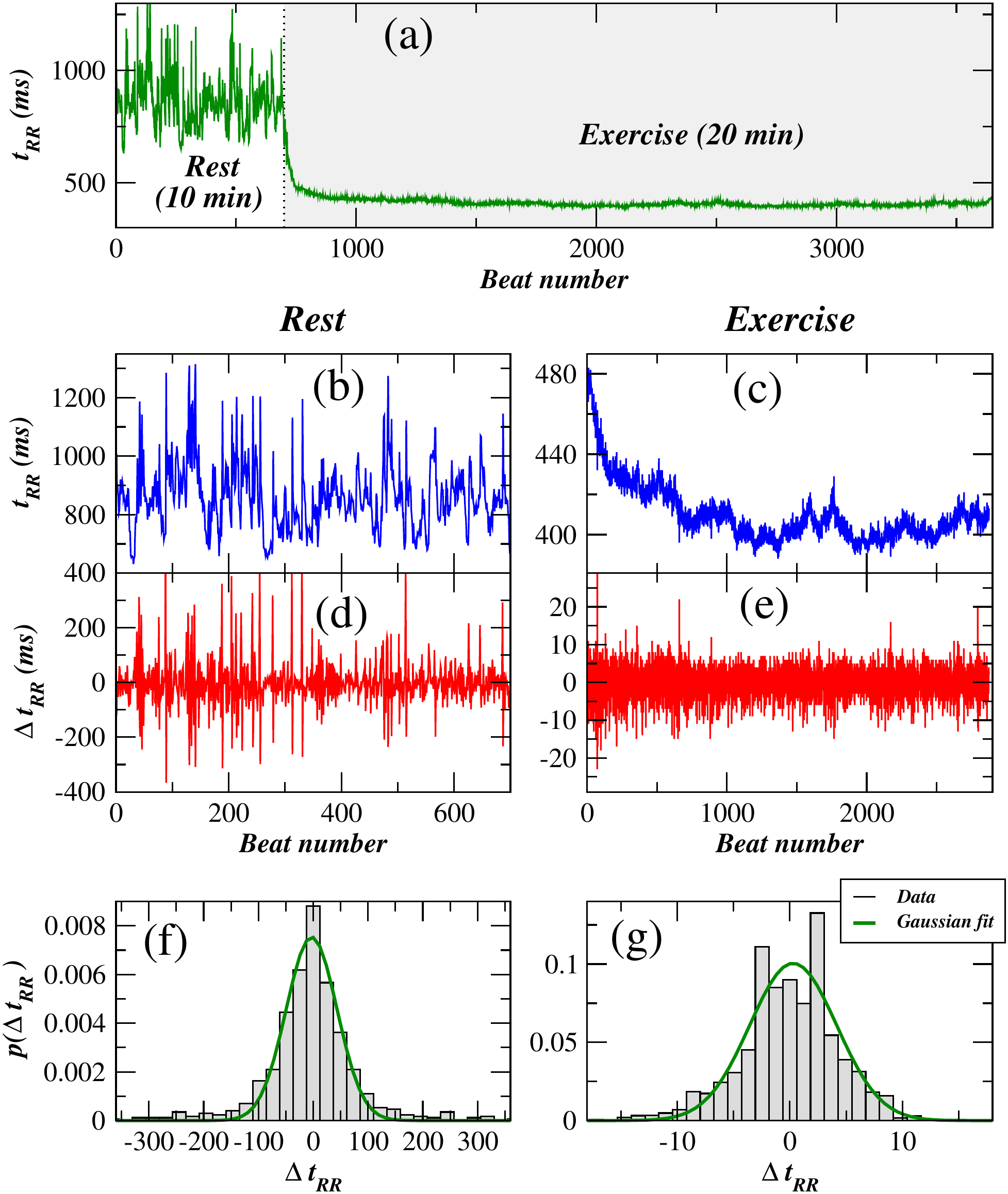}
 \caption{(Color online) Record of interbeat intervals $t_{RR}$ during rest and moderate exercise for a semi-professional soccer player. (a) Full record of 10 minutes resting in supine position on the soccer field pus 20 minutes running at moderate pace. (b)(c) Separate records for rest and exercise. (d) (e) Series of  $\Delta t_{RR}$ for rest and exercise. (f)(g) Distributions of $\Delta t_{RR}$. For comparison it has been included the best fit to a Gaussian distribution (thick line).}
 \label{fig_ej_natural_series}
 \end{figure}

\section{Example of natural signals: Heart rate during rest and exercise}

\label{section_natural_signals}

  {Since the pioneering works  \cite{Peng93}, much attention has been paid to the study of correlations in time series of interbeat intervals, i.e. series of times between consecutive heart beats $\lbrace t_{RR,i} \rbrace$, also known as  $RR$ time series.  In fact, the correlations in such series have} been revealed as a powerful tool to evaluate alterations due to disease or aging \cite{Goldberger02,Plamen1996},  discriminate between physiological states \cite{PlamenSleep1999}  and assess the state of fitness \cite{Aubert03,Dong16}. In most cases,  studies are limited to linear correlations (power-spectrum, autocorrelation function, DFA, etc.) but nonlinear correlations are indeed present in $RR$ time series \cite{Baillie09} and are supposed to play an important role in heart dynamics as their reduction or absence has been related to aging and certain pathological conditions \cite{Plamen1999,Yosi_PRL}.  It is worth mentioning that frequently the correlations of the data are supposed to scale as power laws.

Regarding the heart rate during exercise, it is well known that heartbeat dynamics can change dramatically with physical activity. The most evident changes are the abrupt increase in the heart-rate (i.e. reduction of the mean $RR$ intervals) and the reduction of the heart rate variability (HRV), i.e. the variance of the $RR$ times series \cite{Sarmiento13}. 
In addition to these features that can be observed by direct inspection of raw $RR$ time series (Fig. \ref{fig_ej_natural_series}.a), it has also been found that exercise modifies the distribution of the power spectrum by reducing the low frequency components \cite{Sarmiento13,Sandercock06,Anosov00} and introducing very high frequencies related to the respiration rate \cite{Lewis10}, decreases the sample entropy \cite{Platisa08}. Also, the linear correlations measured by the short scale DFA exponent ($\alpha_1$)  are not only reduced with exercise \cite{Karasik02,Platisa08} but also  can be correlated with the intensity of the exercise \cite{Hautala03}. Nevertheless it is fair to say that the opposite result can also be also found in the literature  \cite{Tulppo2001}.  In summary, despite this last contradiction,  the general agreement is that in a wide sense the complexity of the $RR$ time series is reduced during exercise and that this effect is related to the breakdown of the equilibrium between the two branches of the autonomic nervous system due to the withdrawal of parasympathetic tone and/or the activation of sympathetic activity (see \cite{Sandercock06,Lewis10} for reviews). 

Here we hypothesize that this reduction in complexity should also be reflected in the lost of nonlinearity in the heart dynamics during exercise. In particular, we focus  ourselves on short scales because it has been reported that in this range ($\ell < 11 $ beats) linear correlations seem to be clearly affected  by the intensity of the exercise and because, in practice, the  typical length of the records at rest is rarely longer than 10-15  minutes (500-1000 beats) to avoid excessive interferences with the training sessions, thus preventing from accurate evaluation of autocorrelation functions at long distances. 

We analyze records during rest and moderate exercise from 10 semi-professional soccer players all of them healthy males (age $23.8 \pm 2.9$ yr) without any prior history of cardiovascular disease. Each record includes two stages: (i) 10 minutes of normal wake rest condition, laying in supine position on the soccer field (ii) followed by 20 minutes of moderate running, i.e. at typical warming-up pace (Fig. \ref{fig_ej_natural_series}.a). Heart rate was monitored beat-by-beat using a Polar S810i $RR$ cardiotachometer (Polar Electro, Oy, Finland) \cite{validez_polar2}.

As $RR$ time series are typically non-stationary, especially during exercise (Figs. \ref{fig_ej_natural_series}.b and c), it is a common practice to analyze the series of its increments:
 \begin{equation}
  \Delta t_{RR,i} = t_{RR,{i+1}}-t_{RR,i}
\end{equation}
which are quite stationary, at least in weak-sense (Figs. \ref{fig_ej_natural_series}.d and e). Following the notation introduced in Sec. \ref{section_cmag} $\{y_i\}$ would be the series of $RR$ intervals while $\{x_i\}$ would be the series of interbeat intervals increments ($\Delta t_{RR}$).

The distributions of $\Delta t_{RR}$ are fairly symmetric, although they are not exactly Gaussian but  Levy-stable distributions with tails decaying slower than in the Gaussian case \cite{Peng93} (Figs. \ref{fig_ej_natural_series}.f and g). For this reason, prior to the analysis we convert the distribution of the data to a standard normal distribution by means of the transformation:
\begin{equation}
   x'=\Phi^{-1}\left[ F(x) \right]
\end{equation}
where $F(\cdot)$ is the cumulative distribution of the original $\Delta t_{RR}$ data and $\Phi(\cdot)$ is the cumulative standard normal distribution ${\cal N}(0,1)$. We have observed that this transformation practically does not modify the linear correlations $\C$ (not shown).
 
 For each subject  we compute the autocorrelation function of the series of increments $\C(\ell)$ and of the magnitude series $\Cabs(\ell)$ for both rest and exercise records.
 
   \begin{figure}
   \centering
   \includegraphics[width=7.5cm]{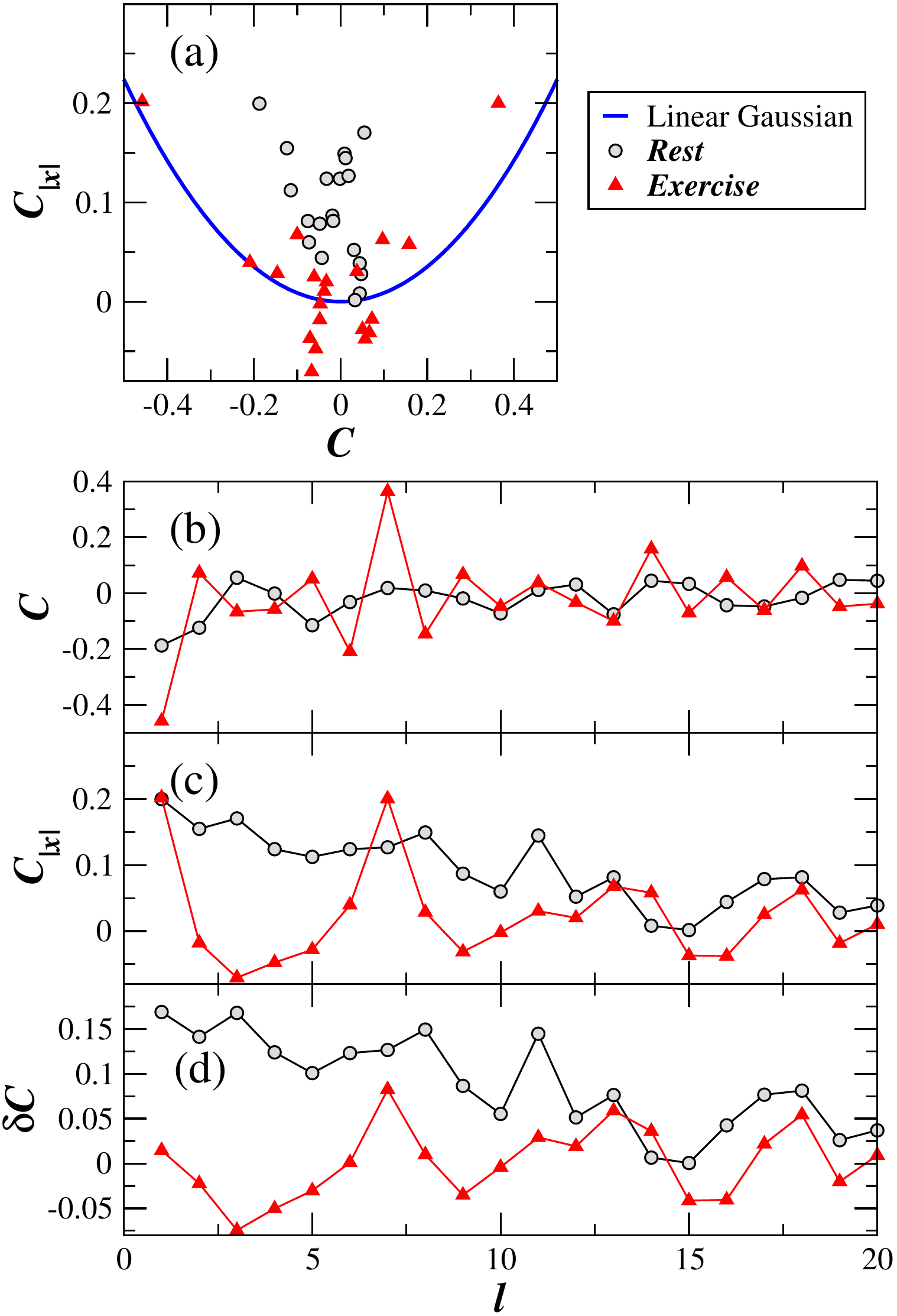}
   \caption{(Color online) Autocorrelation $\C$ and autocorrelation of the magnitude $\Cabs$ for the series of $\Delta t_{RR}$ during rest and exercise shown in Fig.  \ref{fig_ej_natural_series}. (a) $\Cabs$ vs. $\C$ during rest (circles) and exercise (triangles). The thick line corresponds to the theoretical expectation for a linear Gaussian noise (\ref{cmag2}). (b) Autocorrelation $\C(\ell)$ as a function of the lag $\ell$ during rest and exercise. (c) Autocorrelation of magnitude series $\Cabs(\ell)$ as a function of the lag $\ell$ during rest and exercise. (d) Difference between $\Cabs$ and the theoretical expectation for a linear Gaussian noise given $\C$ (see text) as a function of the lag $\ell$ during rest and exercise.} 
   \label{fig_ej_corr_natural}
   \end{figure}

In Fig.~\ref{fig_ej_corr_natural} we show the results for one of the subjects for $\ell=1,...,20$.  In general, we observe that $\C$ reaches similar values during rest and exercise or even greater values for the latter (Fig.~\ref{fig_ej_corr_natural}.b) but, on the other hand,  $\Cabs$ is typically greater during rest (Fig.~\ref{fig_ej_corr_natural}.c). In addition, if we inspect carefully Fig.~\ref{fig_ej_corr_natural}.a it is clear that not only the values of $\Cabs$ are greater on average for rest than for exercise but also  the exercise records are closer to the thick line (expectation for a Gaussian linear noise). For this reason, a good measure of nonlinearity is not simply the  autocorrelation in the magnitude $\Cabs$ but its difference with the expectation for a linear Gaussian noise computed using Eq. (\ref{cmag2}) (thick line in  Fig. \ref{fig_ej_corr_natural}.a): 

 \begin{equation}
  \delta C(\ell) = \Cabs(\ell)-C_{\abs{x},{\rm linear}}(\C(\ell))
 \label{deltac}
 \end{equation}
This quantity takes into account not only the value of $\Cabs$ but also its difference with the linear expectation. For example $\Cabs(\ell=1)$ reaches a relatively high value for both, rest an exercise (Fig. \ref{fig_ej_corr_natural}.c) but, once subtracted the linear expectation $\delta C(\ell=1)$ is much higher for rest than for exercise (Fig. \ref{fig_ej_corr_natural}.d).

\begin{figure}
\centering
\includegraphics[width=8cm]{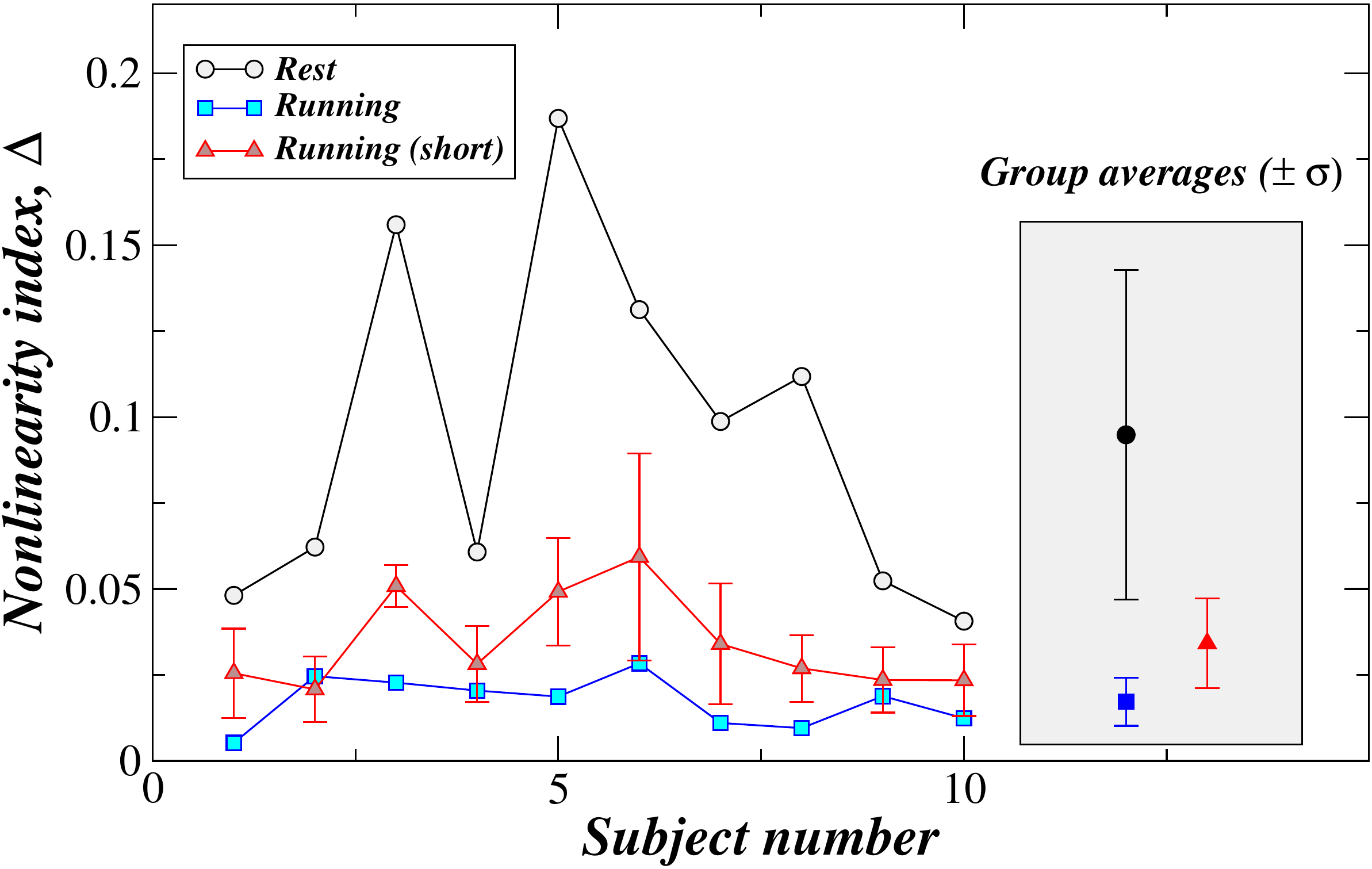} 
\caption{(Color online) Nonlinearity index $\Delta$ ($\ell_{\rm max}=10$) for 10 semiprofessional soccer players, all males with age $23.8 \pm 2.9$ yr. Circles: records of 10 minutes of normal wake rest condition, laying in supine position on the soccer field. Squares: 20 minutes of running at typical warming-up pace. Triangles: average of $\Delta$ over an ensemble of sub-series of the running record with the same size of the corresponding rest record (see text). Error bars indicate $\pm$ standard deviation.}

\label{fig_compara_group}
\end{figure}

In order to obtain a single number to quantify the nonlinearity of a signal we propose here the sum  of the squares of the curve  $\delta C(\ell)$:
\begin{equation}
  \Delta=\sum\limits_{\ell=1}^{\ell_{\rm max}} \delta C(\ell)^2
  \label{index}
\end{equation}
In particular, as we are interested in the short scale correlations, and following most of the authors in the bibliography, we adopt $\ell_{\rm max} = 10$.
We obtain that our nonlinearity index $\Delta$ is clearly higher during rest than during exercise (Fig. \ref{fig_compara_group}). For each individual subject  $\Delta$ is higher for his record during rest than for his corresponding record during exercise and also the group averages are clearly different for rest and exercise ($p< 10^{-7}$). 
Nevertheless, we have to take into account that when dealing with relatively short records, comparisons between series of different length can lead to spurious results due to finite size effects. Here, we have that the records during exercise are two times longer than those during exercise; in addition due to the fact that HR increases with physical activity, the records during exercise are 4-5 times longer in number of beats. For this reason, we check the validity of our findings by comparing our records during rest with records of the same number of beats during exercise: consider a subject with a $N_r$ beats record during rest and a corresponding record during exercise of length $N_e > N_r$ beats and let be $n=\lfloor N_e/N_r \rfloor$. We extract $n$ non-overlapping windows from the exercise record starting from left to right and another $n$ non-overlapping windows from right to left (in order to use all available data). For all these $2n$ sub-series we compute $\Delta$ and average for each subject. Results are shown in Fig. \ref{fig_compara_group} (red triangles). Although now the differences between rest and exercise are a bit smaller, all the values of $\Delta$ for rest are above the corresponding values for exercise (including the error bars) and the difference between group averages is still statistically significant ($p=3\times 10^{-4}$ ).

\section{Conclusions}

\label{conclusions}

We have obtained analytically the expression of the autocorrelation of the magnitude series $\Cabs$ of a linear Gaussian noise as a function of its autocorrelation $\C$ as well as several analytical relations involving $\Cabs$, $\C$ and the autocorrelation of the sign series $\Cs$. 
These expressions are useful to study the nonlinear properties of artificial series obtained by models as well as natural series with the great advantage that our approach does not make any prior assumption about the scaling or functional form of the autocorrelation functions. Indeed, the nonlinearity index proposed in section \ref{section_natural_signals} has the advantage that can be evaluated on relatively small samples and does not require scaling in the autocorrelation function.

In particular, we study the nonlinear properties of a Gaussian model designed to produce series with multifractal properties and show that this model generates nonlinear signals for all the values of the parameters even for those leading to monofractal behavior. This means that, although multifractality seems to imply nonlinearity, the reverse is not always true. 

We also analyze natural time series. Specifically, we have shown that the heart-beat records during rest show higher nonlinearities than the records of the same subject during moderate exercise. This behavior is also achieved on average for the analyzed set of 10 semiprofessional soccer players. With this result we show that the nonlinear properties of the heart-beat dynamics is yet another feature supporting that the complexity of the heart-beat is reduced during exercise. 
It is also worth mentioning that our nonlinearity index is sensible to moderate exercise. This means that it could probably be applied to the study of nonlinear properties during exercise at different levels of intensity and thus, it could be of interest to study the changes in the balance between sympathetic and parasympathetic nervous systems during exercise.

\section{Acknowledgments}
This work is partially supported by grants: FQM-7964 and FQM-362 from the Spanish Junta de Andaluc\'{\i}a. P.B. thanks Oleg Usatenko for helpful discussions.


\appendix

\section{Autocovariance of the magnitudes of two linearly correlated Gaussian variables}

\label{append_mod}
Consider two random variables  $\lbrace X,Y  \rbrace$, both with zero mean and unit standard deviation and following the bivariate Gaussian distribution \cite{Multivariate}:

\begin{eqnarray}
  \rho(x,y)  &\equiv& {\rm Prob} \left\{X=x,Y=y \right\}   \label{bivdens} \\ \nonumber
  &=&  \frac{1}{2\pi\sqrt{1-K^2}} \exp \left [ -\frac{x^2+y^2-2Kxy}{2\left( 1- K^2 \right)} \right ]
\end{eqnarray}
where $K=\ave{xy}$ is the covariance of variables $X$ and $Y$, which also coincides with their correlation taking into account that both of them have zero mean and unit standard deviation. Note that from (\ref{bivdens}) it follows that $K=0$ if and only if $\lbrace X,Y \rbrace$ are independent, i.e. they have only linear correlations.

The covariance of $\abs{X}$ and $\abs{Y}$ is given by:
\begin{eqnarray}
 K_{\rm mag} &\equiv& \ave{\abs{x}\cdot\abs{y}} -\ave{\abs{x}}\ave{\abs{y}} \\ \nonumber 
   &=& \intinfinf\!\abs{x} dx \intinfinf\!\abs{y} dy\,\rho(x,y)- \frac{2}{\pi}
   \label{magnitude_covariance}
\end{eqnarray}
where we have used that $\ave{\abs{x}}=\ave{\abs{y}}=\sqrt{2/\pi}$.

Now, changing the integration variables $\xi=x/\Ak$ and $\varphi=y/\Ak$, where $\Ak\equiv\sqrt{1-K^2}$, we obtain:
\begin{eqnarray}
  K_{\rm mag}&=&\frac{2\Ak^3}{\pi} \intinf\! \xi d\xi\, \e{-\frac{\xi^2}{2}} \times  \label{kabs1} \\ \nonumber 
      &\times& \intinf\! \varphi d\varphi\, \e{-\frac{\varphi^2}{2}} \cosh\left( K \xi \varphi \right) - \frac{2}{\pi}  
 \end{eqnarray}
 
 The integral over $\varphi$ can be written as
  \begin{eqnarray}
  \intinf\! &\varphi& d\varphi\, \e{-\frac{\varphi^2}{2}} \cosh\left( K \xi \varphi \right)     \label{kabs2} \\ \nonumber
   &=& \frac{1}{\xi}\frac{\partial}{\partial K} \left[ \intinf\! d\varphi\, \e{-\frac{\varphi^2}{2}} \sinh\left( K \xi \varphi \right)  \right] \\ \nonumber
   & =& \frac{1}{\xi}\frac{\partial}{\partial K} \left[ \modmed \e{\frac{K^2 \xi^2}{2}} {\rm erf} \left( \frac{K \xi}{\sqrt{2}} \right) \right] \\ \nonumber 
   &=& \modmed K \xi \e{\frac{K^2 \xi^2}{2}}{\rm erf} \left( \frac{K \xi}{\sqrt{2}} \right)+1,
  \end{eqnarray}
  where we have used the identity \cite{Gradshteyn}: 
   \begin{eqnarray}
    \intinf d\varphi \, e^{-b \varphi^2} \cosh (a\varphi) =\frac{1}{2}\sqrt{\frac{\pi}{b}}\e{\frac{a^2}{4b}}
  \end{eqnarray}
  with $a=K \xi $, $b=1/2$ and the fact that
  \begin{eqnarray}
     \frac{d}{dx} {\rm erf}(x)=\frac{2}{\sqrt{\pi}}e^{-x^2}.
  \end{eqnarray}
 
  Replacing (\ref{kabs2}) in (\ref{kabs1})
  \begin{eqnarray}
    K_{\rm mag} &=& \medmod K\Ak^3 \intinf\! \xi^2 d\xi\, \e{-\frac{\Ak^2 \xi^2}{2}} {\rm erf}\left(\frac{K \xi}{\sqrt{2}}\right) \nonumber  \\ &+& \frac{2\Ak^3}{\pi} \intinf\! \xi d\xi\, \e{-\frac{\xi^2}{2}} - \frac{2}{\pi}
  \end{eqnarray}
   and using the identity \cite{Ng}
  \begin{eqnarray}
   \intinf\! &\xi^2&\,d\xi \e{-b^2 \xi^2} {\rm erf}(a \xi) \\ \nonumber  &=& \frac{\sqrt{\pi}}{4 b^3}{\rm sign}(a)-\frac{1}{2\sqrt{\pi}} \left[\frac{1}{b^3} \arctan \left(\frac{b}{a}\right) -\frac{a}{b^2(a^2+b^2)} \right]\;\;\;
  \end{eqnarray}
  with $a=\frac{K}{\sqrt{2}}$ and  $b=\frac{\Ak}{\sqrt{2}}$,  we get:
  \begin{eqnarray}
   K_{\rm mag} &=&
   \abs{K} + \frac{2}{\pi} \left[ K^2\Ak + \Ak^3 - K\arctan\left( \frac{\Ak}{K} \right) -1 \right] \\ \nonumber &=& \abs{K} + \frac{2}{\pi} \left[ \sqrt{1-K^2} - K\arctan\left( \frac{\sqrt{1-K^2}}{K} \right) -1 \right].
 \end{eqnarray}
 
 Finally, after some trigonometric manipulation:
 \begin{eqnarray}
 K_{\rm mag}= \frac{2}{\pi} \left( \sqrt{1-K^2} + K\arcsin {K} -1 \right).
  \label{cov_mag}
 \end{eqnarray}

%

\section{Autocovariance of the squares of two linearly correlated Gaussian variables}

\label{append_square}

Considering again, as in Appendix \ref{append_mod}, two Gaussian variables $\lbrace X,Y \rbrace$ following the bivariate Gaussian distribution (\ref{bivdens}), the autocovariance of their squares is given by:
\begin{eqnarray}
K_{\rm sq} &=&  \ave{x^2 \cdot y^2} -\ave{x^2}\ave{y^2} \\ \nonumber 
&=& \intinfinf x^2 dx \intinfinf y^2 dy \; \rho(x,y) - 1 \nonumber \\ \nonumber
&=& \frac{1}{2\pi \Ak}\intinfinf x^2 y^2 \exp \left [ -\frac{x^2+y^2-2Kxy}{2\Ak^2} \right ] -1 ,
\end{eqnarray}
where we have used that $\ave{x^2}=\ave{y^2}=1$. Now, changing the integration variables $\xi=x/\Ak$ and $\varphi=y/\Ak$ we obtain:

\begin{eqnarray}
K_{\rm sq} &=& \frac{\Ak^5}{2\pi}\intinfinf \xi^2 d\xi \exp\left(-\frac{\xi^2}{2}\right) \times \\ \nonumber
&\times& \underbrace{\intinfinf \varphi^2 d\varphi \exp\left(-\frac{\varphi^2}{2}\right) \exp\left(-K\varphi\xi\right)} -1 \nonumber \\ 
&\;&{\sqrt{2\pi} e^{\frac{K^2\xi^2}{2}}\left(1 + K^2\xi^2\right)}
\end{eqnarray}
Taking into account that $\Ak^2=1-K^2$:
\begin{eqnarray}
 & K_{\rm sq} &= \frac{\Ak^5}{\sqrt{2\pi}} \left[\int\limits_{-\infty}^{\infty}\xi^2 d\xi e^{-\frac{\Ak^2\xi^2}{2}} + K^2 \int\limits_{-\infty}^{\infty} \xi^4 e^{-\frac{\Ak^2\xi^2}{2}} \right] -1 \nonumber \\ 
 &=& \frac{\Ak^2}{\sqrt{2\pi}} \int\limits_{-\infty}^{\infty} x^2 dx e^{-\frac{x^2}{2}}+\frac{K^2}{\sqrt{2\pi}}\int\limits_{-\infty}^{\infty} x^4 dx e^{-\frac{x^2}{2}}-1,
\end{eqnarray}
and finally:
\begin{eqnarray}
 K_{\rm sq} = 2 K^2
 \label{cov_sq}
\end{eqnarray}

\twocolumngrid


\begin{thebibliography}{4}
	
	\bibitem{Yosi_Geo} Y. Ashkenazy, et al.: Nonlinearity and multifractality of climate change in the past 420,000 years. Geophys. Res. Lett. 30(22), 2146 (2003).
	
	\bibitem{Schreiber2000} T. Schreiber \& A. Schmitz: Surrogate time series. Physica D 142, 346â€“382 (2000).
	
	\bibitem{Yosi_PRL} Y. Ashkenazy, et al.: Magnitude and Sign Correlations in Heartbeat Fluctuations. Phys. Rev. Lett. 86, 1900-1903 (2001).
	
	\bibitem{allegrini} P. Allegrini, M. Barbi, P. Grigolini, B. J. West: Dynamical model for DNA sequences, Phys. Rev. E. 52, 5281 (1995).
	
	\bibitem{rangarajan} G. Rangarajan \& M. Ding: Integrated approach to the assessment of long range correlation in time series data. Phys. Rev. E 61, 4991 (2000).
	
	\bibitem{fluid} L. Zhu et al.: Magnitude and sign correlations in conductance fluctuations of horizontal oil water two-phase flow. J. Phys. Conf. Ser. 364, 012067 (2012).
	
	\bibitem{geo01} I. Bartos and I.M. J\'anosi: Nonlinear correlations of daily temperature records over land. Nonlin. Processes Geophys. 13, 571-576 (2006).
	
	\bibitem{geo02} Q. Li, Z. Fu, N. Yuan and F. Xie: Effects of non-stationarity on the magnitude and sign scaling in the multi-scale vertical velocity increment. Physica A. 410, 9-16 (2014).
	
	\bibitem{economy} Y.H. Liu, et al.: Statistical properties of the volatility of price fluctuations. Phys. Rev. E. 60, 1390-1400 (199
	
	\bibitem{Yosi_volatility} T. Kalisky, Y. Ashkenazy and S. Havlin: Volatility of linear and nonlinear time series. Phys. Rev. E 72, 011913 (2005).
	
	\bibitem{manolo} M. G\'omez-Extremera, P. Carpena, P.Ch. Ivanov and P.A. Bernaola-Galv\'an: Magnitude and sign of long-range correlated time series: Decomposition and surrogate signal generation. Phys. Rev. E 93, 042201 (2016). 
	
	\bibitem{Plamen1999} P. Ch. Ivanov, et al.: Multifractality in human heartbeat dynamics. 
	Nature  {\bf 399}(6735),   461-465 (1999).
	
	\bibitem{Pena09} M. A. Pe\~na, J. C. Echeverr\'{\i}a, M.T. Garc\'{\i}a and R. Gonz\'alez-Camarena: Applying fractal analysis to short sets of heart rate variability data. Med. Biol. Eng. Comput (2009) {\bf 47}, 709-717 (2009).
	
	\bibitem{Holl15} M. H\"oll \& H. Kantz: The fluctuation function of the detrended fluctuation analysis â€“ investigation on the AR(1) process. Eur. Phys. J. B {\bf 88}, 126 (2015).
	
	\bibitem{Pandelis09} P. Perikakis, et al.: Breathing frequency bias in fractal analysis of heart rate variability. Biological Psychology 82, 82â€“--88 (2009).
	
	\bibitem{Sandercock06} G.R.H. Sandercock \& D.A. Brodie: The use of heart rate variability measures to assess autonomic control during exercise. Scand. J. Med. Sci. Sports {\bf 16} 302-313 (2006).
	
	\bibitem{CarpenaDFA} P. Carpena, M. G\'omez-Extremera, C. Carretero-Campos, P. Bernaola-Galv\'an and A.V. Coronado: Spurious results of Fluctuation Analysis techniques in magnitude and sign correlations. Entropy, 19, 261 (2017).
	
	\bibitem{multifractal} J.W Kantelhardt  et al. Multifractal detrended fluctuation analysis of nonstationary time series. Physica A 316, 87 (2002).
	
	\bibitem{Makse} H. A. Makse, S. Havlin, M. Schwartz and H. E. Stanley: Method for generating long-range correlations for large systems. Phys. Rev. E 53, 5445 (1996).
	
	\bibitem{Beran} J. Beran, {\it Statistics for Log-memory Processes} (Chapman \& Hall, New York, 1994).
	
	\bibitem{Peng94} C.-K.~Peng, S.V.~Buldyrev, S.~Havlin, M.~Simons, H.E.~Stanley, and A.L.~Goldberger: Mosaic organization of DNA nucleotides. Phys. Rev. E 49, 1685-1689 (1994).
	
	
	
	\bibitem{Conchita12} C. Carretero-Campos, P. Bernaola-Galv\'an, P. Ch. Ivanov and P. Carpena: Phase transitions in the first-passage time of scale-invariant correlated processes. Phys. Rev. E {\bf 85}, 011130 (2012).
	
	\bibitem{signum_corr} S. S. Apostolov, F. M. Izrailev, N. M. Makarov, Z. A. Mayzelis, S. S. Melnyk, O. V. Usatenko: The Signum function method for the generation of correlated dichotomic chains. J. Phys. A: Math. Theor. {\bf 41}, 17501 (2008).
	
	
	
	\bibitem{Parisi85} G. Parisi \& U. Frisch: On the singularity structure of fully-developed turbulence, in Turbulence and predictability in geophysical fluid dynamics, Proc. Int. School E. Fermi, edited by M. Ghil et al., North Holland (1985).
	
	\bibitem{Badin16} G. Badin \& D. I. V. Domeisen: Nonlinear stratospheric variability: multifractal de-trended fluctuation
	analysis and singularity spectra. Proc. R. Soc. A. 472: 20150864 (2016).
	scand j med sci sports
	
	\bibitem{Tang2015} L. Tang, et al.: Complexity testing techniques for time series data:
	A comprehensive literature review. Chaos, Solitons \& Fractals 81, 117 (2015).
	
	
	\bibitem{Peng93} C. K. Peng, et al.: Long-range anti-correlations and non-Gaussian behavior of the heartbeat. Phys. Rev. Lett. {\bf 70}, 1343 (1993).
	
	
	\bibitem{Goldberger02} A. L. Goldberger, et al.: Fractal dynamics in physiology: Alterations with disease and aging.  Proc. Natl. Acad. Sci. USA {\bf 99}, 2466-2472 (2002).
	
	
	\bibitem{Plamen1996} P. Ch. Ivanov et. al.: Scaling behaviour of heartbeat intervals obtained by wavelet-based time-series analysis. Nature {\bf 383}(6598), 323-326 (1996).
	
	
	
	\bibitem{PlamenSleep1999} P. Ch. Ivanov, et al.: Sleep-wake differences in scaling behavior of the human heartbeat: Analysis of terrestrial and long-term space flight data. Europhys. Lett. {\bf 48}(5), 594-600 (1999). 
	
	\bibitem{Aubert03} A.E. Aubert, B. Seps and F. Beckers: Heart	Rate	Variability	in	Athletes. Sports Med.  {\bf 33}(12), 889-919 (2003).
	
	
	\bibitem{Dong16} J.G. Dong: The role of heart rate variability in sports physiology (Review). Experimental and Therapeutic Medicine {\bf 11}, 1531-1536 (2016).
	
	
	\bibitem{Baillie09} R. T. Baillie, A. A. Cecen and C. Erakl: Normal heartbeat series are nonchaotic, nonlinear, and multifractal: New evidence from semiparametric and parametric tests. Chaos {\bf 19}, 028503 (2009).
	
	
	
	
	\bibitem{Sarmiento13} S. Sarmiento et al.: Heart rate variability during high-intensity exercise. J Syst Sci Complex {\bf 26} 104-116 (2013).
	
	
	
	
	\bibitem{Anosov00} O. Anosov, A. Patzak and Y. Kononovich: High-frequency oscillations of the heart rate during ramp load reflect the human anaerobic threshold. Eur. J. Appl. Physiol {\bf 83}, 388-394 (2000).
	
	\bibitem{Lewis10} M. J. Lewis  \& A. L. Short: Exercise and cardiac regulation: what can electrocardiographic time series tell us? Scand. J. Med. Sci. Sports {\bf 20} 794-804 (2010).
	
	\bibitem{Platisa08} M.M. Platisa et al.: Complexity of heartbeat interval series in young healthy trained and untrained men. Physiol. Meas. {\bf 29}, 439-450 (2008).
	
	
	\bibitem{Karasik02} R. Karasik, et al.: Correlation differences in heartbeat fluctuations during rest and exercise. Phys Rev. E 66, 062902 (2002).
	
	\bibitem{Hautala03} A.J. Hautala et al.: Short-term correlation properties of R-R interval dynamics at different exercise intensity levels. Clin Physiol. Funct. Imagin {\bf 23}, 215-223 (2003).
	
	
	
	\bibitem{Tulppo2001} M. P. Tulppo et al.: Effects of exercise and passive head-up tilt on fractal an complexity properties of heart rate dynamics. Am J. Physiol. Heart Circ. Physiol. {\bf 280}, H1081-H1087 (2001).
	
	\bibitem{validez_polar2} M. Weippert, et al.: Comparison of three mobile devices for measuring R-R intervals and heart rate variability: Polar S810i, Suunto t6 and an ambulatory ECG system. European Journal of Applied Physiology {\bf 109}(4), 779-786 (2010).
	
	
	\bibitem{Multivariate} R. A. Johnson \&  D.W. Wichern: {\it Applied Multivariate Statistical Analysis}. (6$^{th}$ Ed. Pearson Prentice Hall, New Jersey 2007).
	
	
	
	\bibitem{Gradshteyn} I.S.Gradshteyn \& I.M. Ryzhik: {\it Table of Integrals, Series and Products}. (Academic Press, New York 1980). Page 356.
	
	\bibitem{Ng} E.W. Ng \& M. Geller: A Table of Integrals of the Error Functions. J. Res. Nat. Bur. Standards Sect. B {\bf 73B}, 1-20 (1969).
	
	
	
	
	
	
	
	
	
	
	
	
	
	
	
	
	
	
	
	
	
	
	
	
	
	
	
	
	
	
	
	
	
	
	
	
	
	
	
	
	
	
	
\end{thebibliography}
\end{document}